\documentclass[twocolumn,pra,showpacs,superscriptaddress]{revtex4}
\usepackage{amssymb}
\usepackage{amsmath}
\usepackage{graphicx}
\usepackage{amsfonts}
\usepackage{color}

\providecommand{\U}[1]{\protect\rule{.1in}{.1in}}
\newtheorem{lemma}{Lemma}
\newtheorem{theorem}{Theorem}

\begin{document}

\title{\textbf{Indirect control with quantum accessor (I): \\
coherent control of multi-level system via qubit chain }}
\author{H. C. \surname{Fu}}
\email{hcfu@szu.edu.cn}
\affiliation{School of Physics, Shenzhen University, Shenzhen 518060, China}
\author{Hui \surname{Dong}}
\affiliation{Institute of Theoretical Physics, Chinese Academy of Sciences, Beijing,
100080,China}
\author{X. F. \surname{Liu}}
\affiliation{Department of Mathematics, Beijing University, Beijing 100871,China}
\author{C. P. \surname{Sun}}
\email{suncp@itp.ac.cn}
\homepage{http://www.itp.ac.cn/~suncp}
\affiliation{Institute of Theoretical Physics, Chinese Academy of Sciences, Beijing,
100080,China}
\date{\today }

\begin{abstract}
Indirect controllability of an arbitrary finite dimensional quantum system ($%
N$-dimensional qudit) through a quantum accessor is investigated. Here, The
qudit is coupled to a quantum accessor which is modeled as a fully
controllable spin chain with nearest neighbor (anisotropic) XY-coupling. The
complete controllability of such indirect control system is investigated in
detail. The general approach is applied to the indirect controllability of
two and three dimensional quantum systems. For two and three dimensional
systems, a simpler indirect control scheme is also presented.
\end{abstract}

\maketitle

\section{Introduction}

Quantum control is essentially understood as a coherence preserving
manipulation of a quantum system, which enables a time evolution from an
arbitrary initial state to an arbitrarily given target state \cite%
{book1,book2,book3,lloyd}. Recently quantum control has attracted much
attention due to its intrinsic relation to quantum information processing
algorithms \cite{Tarn}. It has been demonstrated that the universality of
quantum logic gates can be well understood from the viewpoint of quantum
controllability \cite{book-qin}, and the tools of quantum coherent control
may be used to design protocols of quantum computing \cite{qc-qi}.

In connection with the fundamental limit of quantum information processing
in physics, we have developed an indirect scheme for quantum control \cite%
{xue} where the \emph{controller} is a quantum system and the operations of
quantum control are determined by the initial state of the quantum \emph{%
controller}. This scheme has a built-in feedback mechanism impliedly, which
enables the quantum \emph{controller} to probe the status of the controlled
system and then to manipulate its instantaneous time evolution in coherent
process. However, due to the quantum decoherence induced by the quantum
control itself, the quantum controllability is limited by some uncertainty
relations in the designed quantum control process. The key point in this
approach is that the controller itself needs to be well controlled for the
exact preparation of a proper initial state. Now, this approach motivates us
to generally investigate \emph{indirect control} in which the "quantized
controller" (or quantum accessor) interacts with the controlled system
coherently, and a classical external field couples with the quantum accessor
only to fully control the quantum accessor. From physical point of view the
indirect control is undoubtedly meaningful. Actually, in many physical
situations it is very difficult to control the state of quantum system
directly, but it is easy to manipulate the state of quantum accessor and
thus the state of the \emph{system} via their fixed interaction.

Quantum controllability has been well defined \cite{Tarn} and
extensively studied \cite{Tarn-infinite}. For finite-dimensional
quantum system the complete controllability is well established when
the coupling between the controlled system and external classical
fields is under dipole approximation \cite{fu1,fu2}. From these
results we observe that it is not difficult to design a quantum
accessor which can be well controlled to arrive at an expected
initial state. In fact, for the simple case where both the
controlled system and the quantum accessor are spin-$1/2$ particles,
the controllability problem has been investigated most recently
\cite{ind1, ind2} in the spirit of Refs.~\cite{vile,mandi}, which
consider quantum controllability in connection with quantum
measurement. We consider the problem of indirect controllability of an
arbitrary finite dimensional quantum system by coupling it to a
quantum accessor, a fully controllable spin chain with nearest
neighbor (anisotropic) XY-coupling (see Fig.2).

\begin{figure}[tbp]
\includegraphics[bb=100 160 500 640, width=7 cm, clip]{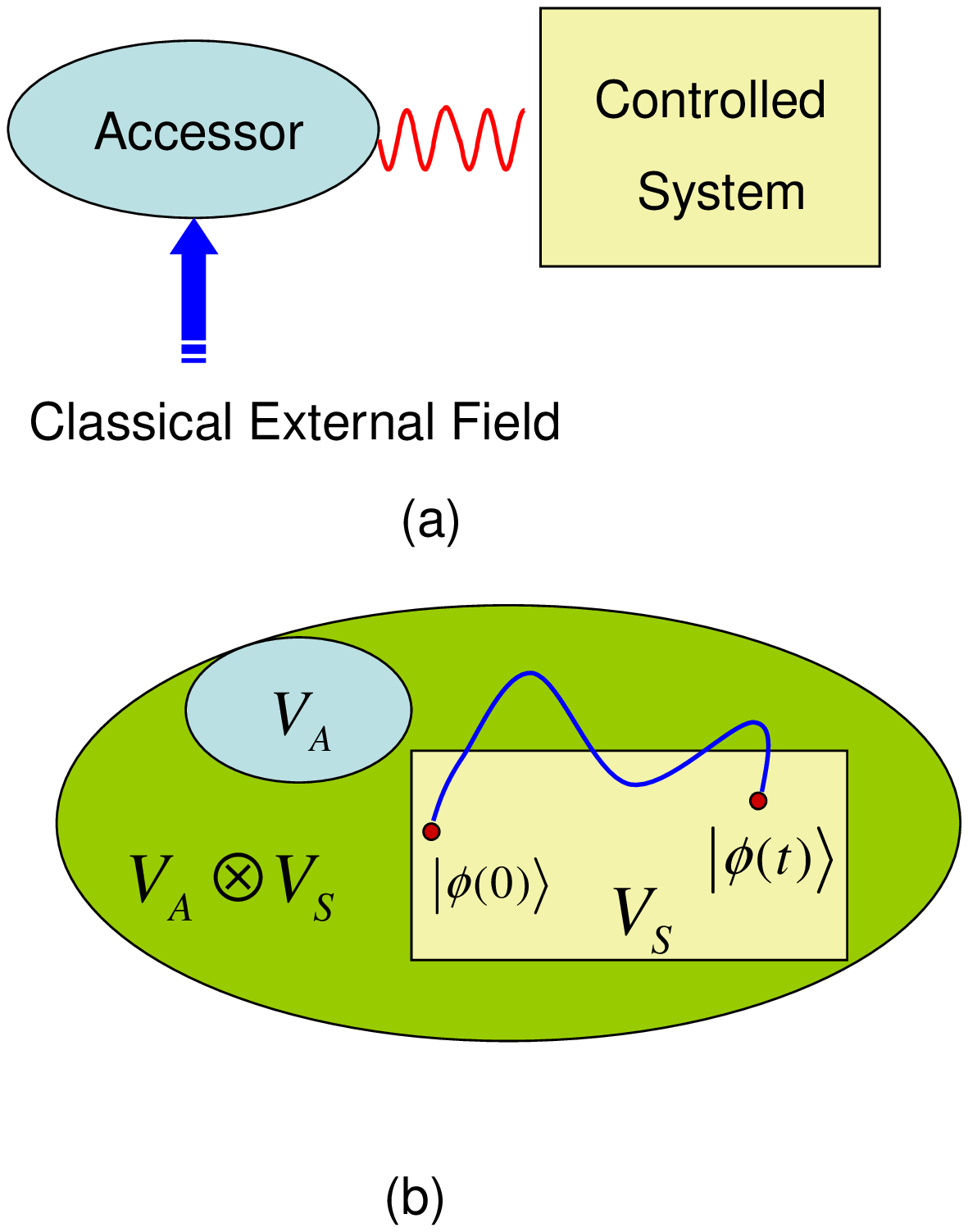}
\caption{Illustration of indirect quantum control: (a) An external field
classically manipulates the quantum accessor and then indirectly controls
the quantum system coupling to the accessor with a fixed interaction. (b)
When each state in the total Hilbert space $V_{S}\otimes V_{A}$ is reachable
under the control via the external classical field acting on the
accessor only, each state in the Hilbert space $V_{S}$ must be reachable. This
enables a complete controllability for the indirect control of the
controlled system }
\end{figure}
In this paper we utilize the Lie algebra method to systematically study the
controllability of the total system formed by the controlled quantum system $%
\mathcal{S}$ and the quantum accessor $\mathcal{A}$ with Hamiltonian $%
H_{0}=H_{S}$ $+H_{A}+H_{SA}$. In the theoretic framework of quantum control,
it is assumed that the time evolution of the total system can be externally
controlled by a family of additional steering fields $\{u_{j}(t)\}$ in a
suitable parameter space through the control Hamiltonian
\begin{equation}
H_{c}=\sum_{j}u_{j}(t)W_{j}(a,s).
\end{equation}%
Here $H_{S}=H_{S}(s)$ ($H_{A}=H_{A}(a)$) is the free Hamiltonian of $%
\mathcal{S}$ ($\mathcal{A}$) of variable $s$ ($a$) defined on the Hilbert
space $V_{S}$ ($V_{A}$) and the coupling Hamiltonian $H_{SA}=H_{SA}(s,a)$
between the system $\mathcal{S}$ and the accessor $\mathcal{A}$ is generally
defined on the space $V_{S}\otimes V_{A}$. The control operators $W_{j}(a,s)$
are usually defined also on $V_{S}\otimes V_{A}.$

Obviously it is rather trivial to consider the controllability of the total
system of $\mathcal{S}$ and $\mathcal{A}$ when $W_{j}(a,s)$ depends on both $%
s$ and $a$ since this is essentially the conventional classical control
problem of the composite quantum system of $\mathcal{S}$ and $\mathcal{A}$.
But it is equally obvious that an important situation will arise if $%
W_{j}(a,s)$ is constrained to the space of accessor, namely, $%
\partial_{s}W_{j}(a,s)=0 $ or $W_{j}(a,s)=W_{j}(a)$. This case is not at all
trivial: it suggests the possibility of controlling the quantum system $%
\mathcal{S}$ through the control of the variables of the quantum accessor.
In fact, this situation is exactly what we will probe in this paper.

We will prove that under some general conditions the control of $\mathcal{A}$
variables can indeed result in a complete quantum control of the whole
system and thus lead to an ideal control of its subsystem, the original
controlled quantum system $\mathcal{S}$. From mathematical point of view, if
the whole system is ergodic in the whole Hilbert space $V_{S}\otimes V_{A}$,
then each state in the subspace $V_{S}$ must be reachable by the subsystem $%
\mathcal{S}$ in the same control process. Here we should point out that a
broad dynamical-algebraic framework has been presented, from different
motivations and approaches, for analyzing the quantum control properties in
terms of the group representation theory \cite{z1,z2}.

In this paper, the first one of our series papers on indirect quantum
control, we shall consider the indirect controllability of arbitrary $N$%
-energy level quantum system (the qubit) $\mathcal{S}$ through an accessor $%
\mathcal{A}$ modeled as the spin chain of XY type with nearest neighbor
coupling. The controlled system $\mathcal{S}$ and the accessor $\mathcal{A}$
are coupled constantly. We control the system $\mathcal{S}$ by controlling
each individual spin of the accessor through a family of external classical
fields. To the end of indirect control of quantum system through accessor we
also apply a constant classical field to excite the system to be controlled.
However, as we will discuss for the case of the 2-dimensional system (see
Eq.\thinspace (\ref{remove})), such constant excitation can be removed by
rotating the controlled system. In the terminology of group theory, this
quantum control problem is cased to the Lie group structure \cite{hum,
Helgason}
\begin{equation}
U(N)_{S}\otimes G_{A}=U(N)_{S}\otimes U(2)_{1}\otimes ...\otimes U(2)_{M}.
\end{equation}

The remaining part of this paper is organized as follows. In section II, we
model the controlled system $\mathcal{S}$ and the accessor $\mathcal{A}$,
and formulate the indirect control system. In Section III, we systematically
investigate the conditions concerning the complete controllability of the
indirect control system, including the coupling between the system and the
accessor. In Sections IV and V, we apply the general approach to
two and three dimensional cases, respectively. Besides, for the two and
three dimensional systems, we will discuss more economical indirect control.
Finally, we make a short summary and some remarks in Section VI.


\section{Indirect quantum control with multi-qubit encoding}

First of all, let us point out that throughout this paper the symbol $i$
stands for the complex number $\sqrt{-1}$.

Let $\mathcal{S}$ be the $N$- level quantum system (or qudit) with energy
levels $|j\rangle $ ($j=1,2,...,n$), described by the Hamiltonian
\begin{equation}
H_{S}=\sum_{j=1}^{N}E_{j}e_{jj}.
\end{equation}%
Here $E_{j}$ is the eigen energy and the projection operator $%
e_{jk}=|j\rangle \langle k|$ stands for the $N\times N$ matrix with the
entries $(e_{jk})_{lm}=\delta _{jl}\delta _{km}$. Without losing generality,
we suppose that the Hamiltonian $H_{S}$ is traceless, namely tr$H_{S}=0$ or $%
\sum_{j=1}^{N}E_{j}=0$. Our aim is to answer the question: can we steer the
system $\mathcal{S}$ from an initial state to a target state through an
intermediate quantum system, the accessor $\mathcal{A}$ and a family of
classical fields which control the accessor $\mathcal{A}$ only?

\begin{figure}[tbp]
\includegraphics[bb=53 288 496 589,width=8 cm, clip]{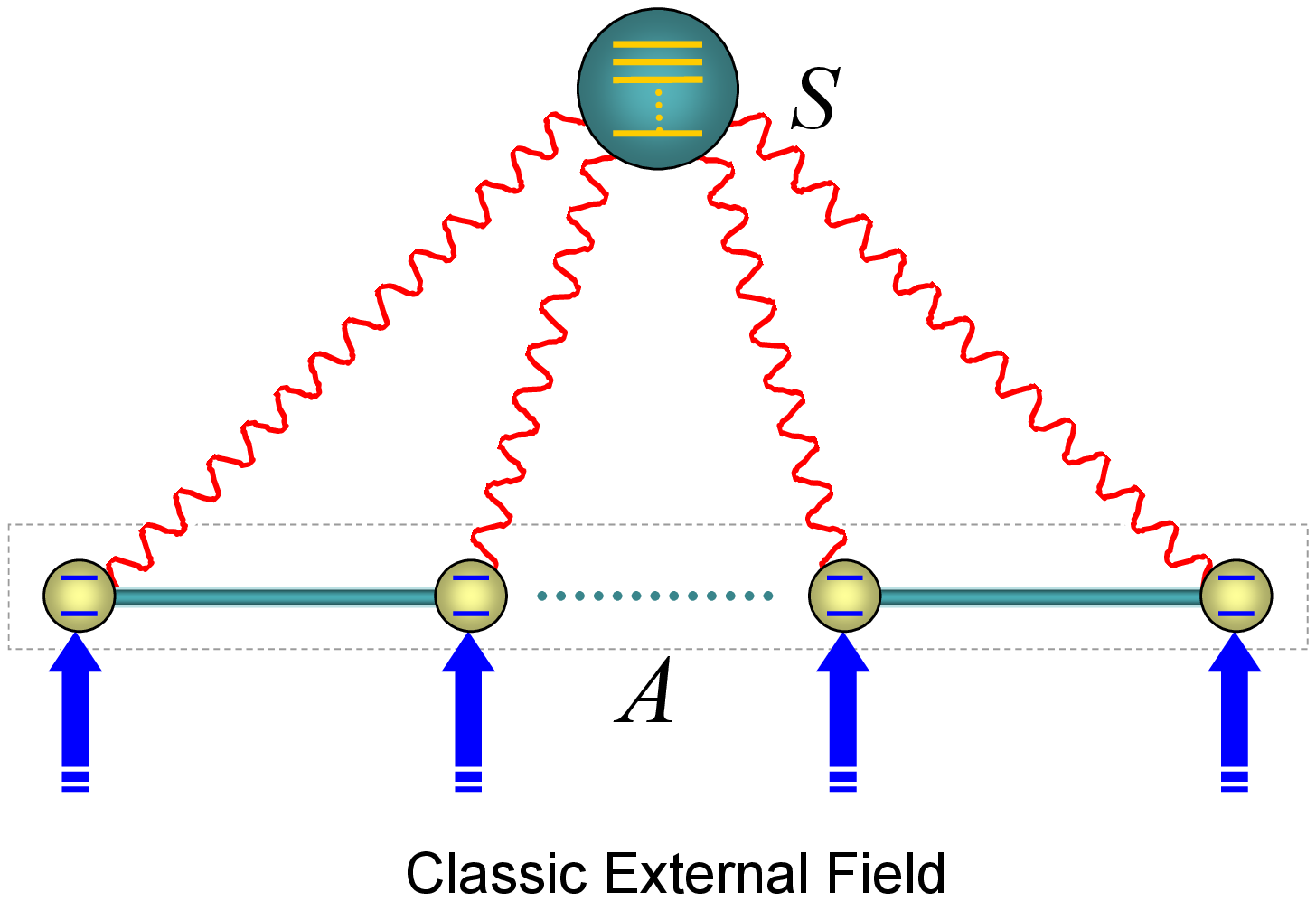}
\caption{The indirect control system consists of a quantum accessor $%
\mathcal{A}$ and a $N$-level controlled system $\mathcal{S}$. Here $M$
qubits coupled through nearest neighbor interaction work as the accessor $%
\mathcal{A}$. We indirectly control the system $\mathcal{S}$ by manipulating
the accessor $\mathcal{A}$ with the classic external field.}
\label{fig:setup}
\end{figure}

Intuitively, we need a high dimensional accessor $\mathcal{A}$ to control a
high dimensional controlled system. We will use a qubit chain to implement
this high dimensional accessor $\mathcal{A}$. Suppose that $\mathcal{A}$
consists of $M$ qubits coupled through nearest neighbor interaction with the
Hamiltonian $H_{A}=H_{A}^{0}+H_{A}^{\prime }:$
\begin{equation}
H_{A}^{0}=\sum_{j=1}^{M}\hbar \omega _{j}\sigma _{z}^{j},\ \ H_{A}^{\prime
}=\sum_{j=1}^{M-1}c_{j}\sigma _{x}^{j}\sigma _{x}^{j+1},  \label{is}
\end{equation}%
where $c_{j}\neq 0$ is the coupling constant of the nearest neighbor
interaction of qubits, $2\hbar \omega _{j}$ is the level spacing of the $j$%
-th qubit, and $\sigma _{\alpha }^{j}$ ($\alpha =x,y,z$; $j=1,2,\cdots ,M$)
is the Pauli's matrix $\sigma _{\alpha }$ of the $j$-th qubit
\begin{equation}
\sigma _{\alpha }^{j}=1\otimes \cdots \otimes 1\otimes \sigma _{\alpha
}\otimes 1\otimes \cdots \otimes 1.
\end{equation}%
The Hamiltonian (\ref{is}) describes the well known Heisenberg model
with nearest neighbor XY-coupling and can be
used to simulate a quantum computer by appropriate coding \cite{isc}. The
setup of control system is schematically illustrated in Fig.\ref{fig:setup}.

To control the system $\mathcal{S}$ through $\mathcal{A}$, $\mathcal{S}$ has
to be coupled to $\mathcal{A}$. We first excite the system $\mathcal{S}$ by
applying a \emph{constant} classical field on the system $\mathcal{S}$ via
the dipole interaction
\begin{equation}
H_{S}^{\prime }=\sum_{j=1}^{N-1}d_{j}x_{j}\otimes 1_{A},
\end{equation}%
where $d_{j}$'s are time-independent real coupling constants, and $x_{j}$'s
are the Hermitian operators defined as $x_{j}=e_{j,j+1}+e_{j+1,j}$. For
later use we define $x_{jk}$, $y_{jk}$ ($1\leq j<k\leq N$)and $h_{j}$ as
follows:
\begin{align}
& x_{jk}=e_{jk}+e_{kj},  \notag \\
& y_{jk}=i(e_{jk}-e_{kj}),  \notag \\
& h_{j}=e_{j,j}-e_{j+1,j+1}.  \label{chebasis}
\end{align}%
Notice that $x_{j}=x_{j,j+1}$ by definition. For this reason, let us define $%
y_{j}=y_{j,j+1}$. We remark here that with the fixed couplings of $\mathcal{S%
}$ to an external field, the Hamiltonian of $\mathcal{S}$ can still be
diagonalized to take the same form as that of $H_{S}$, but the interaction (%
\ref{is}) between $\mathcal{S}$ and $\mathcal{A}$ will then have a
complicated form. The skew-Hermitian operators $ix_{jk}$, $iy_{jk}$ and $%
ih_{j}$ ($1\leq j<k\leq N$)constitute the well-known Chevalley basis of the
Lie algebra su($N$) \cite{hum}. Hereafter we use $1_{S}$ and $1_{A}$ to
denote the identity operator on the Hilbert spaces of the system and the
accessor respectively.

We note that, different from the conventional control problem, here the
interaction $H_{S}^{\prime }$ is time-independent. It seems that the control
scenario considered here is not strictly indirect, since there requires a
constant control field directly coupling all adjacent transitions of the $N$%
-level system. However, the excitation by $d_{j}x_{j}\otimes 1_{A}$ can be
removed by a transformation of the controlled system, which, in effect, will
introduce effective coupling terms to the interaction Hamiltonian
$H_{A}^{\prime }$. The
explicit proof of this point can be found in Section IV where spin 1/2 is
used as an example of the controlled system. We also remark that this
constant control field is introduced only for the convenience of the
presentations of the lemmas and theorems.

In the following discussion, for convenience for $\alpha_j\in\{x,y,z,0\},%
\,j=1,2,\cdots,M$, we use the abbreviation
\begin{equation*}
[\alpha ]=(\alpha_1,\alpha_2,\cdots,\alpha_M),
\end{equation*}
and define
\begin{equation*}
\sigma_{[\alpha]}=\prod_{j=1}^{M}\sigma_{\alpha_{j}}^{j},\,\sigma_0=1.
\end{equation*}

The coupling between the system $\mathcal{S}$ and the accessor $\mathcal{A}$
is generally given as
\begin{equation}
H_{SA}=\sum_{j=1}^{N-1}\sum_{k=1}^{2}\sum_{[\alpha]}g_{[%
\alpha]}^{j(k)}s_{j}^{(k)} \otimes\sigma_{[\alpha]},
\end{equation}
where in the summation over $[\alpha]$ each $\alpha_j$ is restricted to the
set $\{x,y\}$, $s_{j}^{(k)}$ ($1\leq j\leq N-1$, $k=1,2$) denotes either $%
x_{j}$ or $y_{j}$ defined in Eq.\thinspace(\ref{chebasis}):
\begin{equation}
s_{j}^{(k)}=\left\{
\begin{array}{ll}
x_{j}, & \text{when }k=1, \\
y_{j}, & \text{when }k=2,%
\end{array}
\right.
\end{equation}
and $g_{[\alpha]}^{j(k)}$ is the coupling constant. The above coupling is
general for spin-large spin interaction and reduces to the Heisenberg type
coupling when $N=2$.

Then the total system of $\mathcal{S}$ and $\mathcal{A}$ is described by the
Hamiltonian $H_{0}$
\begin{equation}
H_{0}=H_{S}\otimes1_{A}+H_{S}^{\prime}+1_{S}\otimes H_{A}+H_{SA}.
\end{equation}
The central point of our protocol is to control the system $\mathcal{S}$
\emph{indirectly} by controlling the accessor $\mathcal{A}$ using classical
fields. Suppose we can completely control every qubit using two independent
external fields $f_{j}(t)$ and $f_{j}^{\prime}(t)$, $j=1,2,\cdots,M$, which
couple to a qubit in the following way \cite{fu1,fu2}:
\begin{align}
& H_{x}^{j}=1_{S}\otimes\sigma_{x}^{j},  \label{cf1} \\
& H_{y}^{j}=1_{S}\otimes\sigma_{y}^{j}.  \label{cf2}
\end{align}
Then the total Hamiltonian for the indirect control is obtained as
\begin{equation}
H=H_{0}+\sum_{j=1}^{M}\left(
f_{j}(t)H_{x}^{j}+f_{j}^{\prime}(t)H_{y}^{j}\right).  \label{entire}
\end{equation}
In this paper we shall examine under what conditions the control system (\ref%
{entire}) is completely controllable.


\section{Complete controllability of indirect control}


In this section we consider the complete controllability of the system $%
\mathcal{S}$: whether the system $\mathcal{S}$ can be controlled completely
by controlling the accessor $\mathcal{A}$. For this purpose, it is enough to
investigate whether the Lie algebra $\mathcal{L}$ generated by $iH_{0}$, $%
iH_{x}^{j}$ and $iH_{y}^{j}$ is $su$($2^{M}N$), which generates the Lie
group of all the unitary operations on $V_{S}\otimes V_{A}$ through the
single parameter subgroups. If $\mathcal{L}$ is equal to $su$($2^{M}N$), the
system is completely controllable. Otherwise, the system is partly
controllable.

For the skew-hermitian operators
\begin{equation}
iH_{0},\ \ \ \ iH_{x}^{j},\ \ \ \ iH_{y}^{j},\ \ j=1,2,...,M
\label{generators0}
\end{equation}
to generate the Lie algebra $su(2^{M}N)$ some conditions should be
satisfied. This section is mainly devoted to the investigation of such
conditions when $M$ is greater than $2$, the cases with $M=1,2$ being left
to the subsequent sections.

For convenience, we introduce the following notions about conditions on the
system $\mathcal{S}$:

Condition 1. $c_{j}\neq0$ for $j=1,2,\cdots,M-1$;

Condition 2. There exist $2(N-1)\equiv N^{\prime }$ elements $%
[\beta]_1,[\beta]_2,\cdots,[\beta]_{N^{\prime }}$ of the set $\{x,y\}^M$
such that the matrix
\begin{equation}
G=\left[
\begin{array}{cccccc}
g_{[\beta]_1}^{1(1)} & \cdots & g_{[\beta]_1}^{(N-1)(1)} &
g_{[\beta]_1}^{1(2)} & \cdots & g_{[\beta]_1}^{(N-1)(2)} \\
g_{[\beta]_2}^{1(1)} & \cdots & g_{[\beta]_2}^{(N-1)(1)} &
g_{[\beta]_2}^{1(2)} & \cdots & g_{[\beta]_2}^{(N-1)(2)} \\
\cdots & \cdots & \cdots & \cdots & \cdots & \cdots \\
g_{[\beta]_{N^{\prime }}}^{1(1)} & \cdots & g_{[\beta]_{N^{\prime
}}}^{(N-1)(1)} & g_{[\beta]_{N^{\prime }}}^{1(2)} & \cdots &
g_{[\beta]_{N^{\prime }}}^{(N-1)(2)}%
\end{array}
\right]
\end{equation}
is not singular, namely, the determinant of $G$ is nonzero;

Condition 3. The complete controllability conditions on the coupling
constants and the eigen-energy $E_j$, presented in Ref.[10,11].

Notice that Condition 2 implies the restriction $2^{M} \geq2(N-1)$.

\begin{lemma}
Given an arbitrary $[\beta]=(\beta_{1},\beta_{2},\cdots,\beta_{M})\in
\{x,y\}^M$, we have
\begin{align}
& i^{M}\left[ 1_{S}\otimes\sigma_{\beta_{M}}^{M},\left[ 1_{S}\otimes
\sigma_{\beta_{M-1}}^{M-1},\left[ \cdots,\left[ 1_{S}\otimes\sigma
_{\beta_{1}}^{1}, \right.\right.\right.\right.   \notag \\
& \left.\left. i(1_S\otimes H_{A}^{\prime})\right] \cdots \right]  \notag \\
& =\left\{
\begin{array}{ll}
4ic_{1}\delta_{\beta_{1}y}\delta_{\beta_{2}y}(1_{S}\otimes\sigma_{z}^{1}%
\sigma_{z}^{2}), & \text{when }M=2; \\
0 & \text{when }M>2.%
\end{array}
\right.  \label{lemma1}
\end{align}
\end{lemma}

This lemma can be verified directly. We would rather omit the proof.

\begin{lemma}
If $i(1_{S}\otimes\sigma_{x}^{j}\sigma_{x}^{j+1})\in\mathcal{L}$ \textrm{(}$%
j=1,2,\cdots,M-1$\textrm{)}, then for an arbitrary $[\alpha]\in
\{x,y,z,0\}^M $ except $[\alpha]=(0,0,\cdots,0)$ we have $%
i(1_{S}\otimes\sigma_{[\alpha]})\in \mathcal{L}$.
\end{lemma}

\noindent\textbf{Proof.} We first consider the element $i(1_{S}\otimes%
\sigma_{[\alpha]})$ with $\alpha_1=\alpha_2=\cdots=\alpha_M=x$. From (\ref%
{cf1}) and (\ref{cf2}) we have $1_S\otimes\sigma_y^j\in \mathcal{L}$ and
\begin{equation}
-2^{-1}\left[ iH_{x}^{j},iH_{y}^{j}\right] =i(1_S\otimes\sigma_{z}^{j})\in%
\mathcal{L}.  \label{sig3}
\end{equation}

As a result,
\begin{align*}
& 2^{-1}[i(1_S\otimes\sigma_{x}^{2}\sigma_{x}^{3}),\ i(1_S\otimes
\sigma_{z}^{2})]=i(1_S\otimes\sigma_{y}^{2}\sigma_{x}^{3})\in\mathcal{L}, \\
&
-2^{-1}[i(1_S\otimes\sigma_{x}^{1}\sigma_{x}^{2}),\
i(1_S\otimes\sigma_{y}^{2}\sigma_{x}^{3})]= \notag \\
& \ \ \ \ \ \  i(1_S\otimes\sigma_{x}^{1}%
\sigma_{z}^{2}\sigma_{x}^{3})\in\mathcal{L}, \\
&
2^{-1}[i(1_S\otimes\sigma_{x}^{1}\sigma_{z}^{2}\sigma_{x}^{3}),i(1_S\otimes%
\sigma_{y}^{2})]=i(1_S\otimes\sigma_{x}^{1}\sigma_{x}^{2}\sigma_{x}^{3})\in%
\mathcal{L}.
\end{align*}
In the same way we can obtain $i(1_S\otimes\sigma_{x}^{1}\sigma_{x}^{2}%
\sigma_{x}^{3}\sigma_{x}^{4})\in\mathcal{L}$. Now we easily observe that by
repeating this procedure we can prove that
\begin{equation}
i(1_S\otimes\sigma_{x}^{1}\sigma_{x}^{2}\cdots\sigma_{x}^{M})\in\mathcal{L}.
\label{assertion}
\end{equation}

Next, we consider the elements $i(1_{S}\otimes\sigma_{[\alpha]})$ with $%
\alpha_j\in \{x,y,z\}$. It is easy to see that such elements lie in the Lie
algebra generated by $\{i(1_S\otimes\sigma_{x}^{1}\sigma_{x}^{2}\cdots%
\sigma_{x}^{M}), iH_x^j,iH_y^j|j=1,2,\cdots,M\}$, which is a subset of $%
\mathcal{L}$. It then follows that $i(1_{S}\otimes\sigma_{[\alpha]})\in
\mathcal{L}$ for $\alpha_j=x,y,z$.

Finally, we deal with the general element $i(1_{S}\otimes\sigma_{[\alpha]})$
It remains to prove that $i(1_{S}\otimes\sigma_{[\alpha]})\in \mathcal{L}$
for the $\alpha$ with some $\alpha_j^{\prime }s$ being zero. To this end, we
observe that
\begin{equation*}
2^{-1}[i(1_S\otimes\sigma_{x}^{1}\sigma_{x}^{2}),\ i(1_S\otimes
\sigma_{z}^{2})]=i(1_S\otimes\sigma_{x}^{1}\sigma_{y}^{2})\in\mathcal{L},
\end{equation*}
so it follows that

\begin{align*}
& -2^{-1}\left[ i(1_S\otimes\sigma_{x}^{1}\sigma_{x}^{2}\cdots%
\sigma_{x}^{M}),i(1_S\otimes\sigma_{x}^{1}\sigma_{y}^{2})\right] \\
&
=i(1_S\otimes\sigma_{0}^{1}\sigma_{z}^{2}\sigma_{x}^{3}\cdots\sigma_{x}^{M})%
\in\mathcal{L}.
\end{align*}
Now having this element at our disposal, with the help of $iH_x^j$ and $%
iH_y^j$ we can generate in $\mathcal{L}$ all the elements $%
i(1_{S}\otimes\sigma_{[\alpha]})$ with $\alpha_1=0$ and $\alpha_j\in
\{x,y,z\}$, $j\neq 1$. After a moment's thought, one can see that using this
trick we can actually prove that $i(1_{S}\otimes\sigma_{[\alpha]})\in%
\mathcal{L}$ for the $\alpha$ with one $\alpha_j$, not necessarily $\alpha_1$%
, being zero. Finally, along the same way we can proceed further to show
that $i(1_{S}\otimes\sigma_{[\alpha]})\in\mathcal{L}$ for the $\alpha$ with $%
n$ $\alpha_j$'s ($1\leq n< M$) being zero. The lemma is thus proved.

\begin{lemma}
When $M>2$, if Condition 2 is satisfied, then for $j=1,2,\cdots,N-1$ and $%
[\alpha]\ne(0,0,\cdots,0)$ the elements $ix_{j}\otimes\sigma_{[%
\alpha]},iy_{j}\otimes\sigma_{[\alpha]}, ih_{j}\otimes1_A$ lie in $\mathcal{L%
}$.
\end{lemma}

\noindent\textbf{Proof.} We have already known that the elements $%
i(1_S\otimes\sigma_z^j)$($j=1,2,\cdots,M$)are contained in $\mathcal{L}$. So
$i(1_S\otimes H_I^0)$, which is a linear combination of these elements, is
also contained in $\mathcal{L}$. It then follows that $iH_0-i(1_S\otimes
H_I^0)\in \mathcal{L}$, namely,
\begin{equation}
iH_{0}^{\prime}\equiv iH_{S}\otimes1_{A}+iH_{S}^{\prime}+i(1_{S}\otimes
H_{A}^{\prime})+iH_{SA}\in\mathcal{L}.
\end{equation}

Now for $\beta_j\in \{x,y\}$, let us consider the element
\begin{equation}
i^{M}\left[ 1_{S}\otimes\sigma_{\beta_{M}}^{M},\left[ 1_{S}\otimes
\sigma_{\beta_{M-1}}^{M-1},\left[ \cdots,\left[ 1_{S}\otimes\sigma
_{\beta_{1}}^{1},iH_{0}^{\prime}\right] \cdots\right. \right. \right] ,
\label{comm}
\end{equation}
which belongs to $\mathcal{L}$ as $i(1_S\otimes \sigma_{\beta_j}^j)$ belongs
to $\mathcal{L}$ by definition.

Clearly, the term $i(H_{S}\otimes1_{A})+iH_{S}^{\prime}$ in $iH_{0}^{\prime}$
has no nonzero contribution to this element. Moreover, since $M>2$ Lemma 1
tells us that the term $i(1_{S}\otimes H_{A}^{\prime})$ has no nonzero
contribution either.

By straightforward calculation it then follows that
\begin{align*}
& i^{M}\left[ 1\otimes\sigma_{\beta_{M}}^{M},\left[ 1\otimes\sigma
_{\beta_{M-1}}^{M-1},\left[ \cdots,\left[ 1\otimes\sigma_{%
\beta_{1}}^{1},iH_{SA}\right] \cdots\right. \right. \right] \\
& =i(-1)^{M+\Delta}2^{M}\left[ \sum_{j=1}^{N-1}\sum_{k=1}^{2}g_{[\bar {\beta}%
]}^{j(k)}s_{j}^{(k)}\right] \otimes\sigma_{z}^{1}\cdots\ \sigma_{z}^{M}\in%
\mathcal{L},
\end{align*}
where $\bar{\beta}$ is defined as
\begin{equation}
\bar{\beta_{j}}=\left\{
\begin{array}{ll}
x, & \text{{if }}\beta_{j}=y, \\
y, & \text{{if }}\beta_{j}=x,%
\end{array}
\right.  \label{c2}
\end{equation}
and $\Delta$ is the number of $y$ in $\{\beta_{j}|j=1,2,\cdots,M\}$.
Consequently, for each $[\beta]\in \{x,y\}^M$ we have

\begin{equation}
i\left[ \sum_{j=1}^{N-1}\sum_{k=1}^{2}g_{[\bar {\beta}]}^{j(k)}s_{j}^{(k)}%
\right] \otimes(\sigma_{z}^{1}\cdots\sigma_{z}^{M})\in \mathcal{L}.
\label{c3}
\end{equation}
There are altogether $2^M$ such elements. Now Condition 2 guarantees that
from these elements we can choose $2(N-1)$ linearly independent ones. Then
from these linearly independent elements in $\mathcal{L}$ we can derive

\begin{equation}
is_{j}^{(k)}\otimes(\sigma_{z}^{1}\cdots\sigma_{z}^{M})\in\mathcal{L},
\,j=1,2,\cdots,N-1,\,k=1,2  \label{c4}
\end{equation}
by the standard method of linear algebra. Using the same method as that in
the proof of Lemma 2, we can go further to prove that $is_{j}^{(k)}\otimes%
\sigma_{[\alpha]}\in\mathcal{L}$, namely, $ix_j\otimes\sigma_{[\alpha]},iy_j%
\otimes\sigma_{[\alpha]}\in\mathcal{L}$, for $[\alpha]\ne (0,0,\cdots,0)$.
Then the lemma follows directly because we have
\begin{equation*}
(-2)^{-1}\left[ ix_{j}\otimes\sigma_{[\alpha]},iy_{j}\otimes\sigma_{[\alpha]}%
\right] =ih_{j}\otimes1_A.
\end{equation*}

\begin{lemma}
When $M>2$, if Condition 1 and Condition 2 are satisfied, then for $%
[\alpha]\ne(0,0,\cdots,0)$ we have $i1_S\otimes\sigma_{[\alpha]}\in \mathcal{%
L}$.
\end{lemma}

\noindent\textbf{Proof.} We observe that it follows from Lemma 2 that $%
iH_{SA}\in \mathcal{L}$ and $iH_S\otimes 1_A\in \mathcal{L}$. The former is
obvious and the latter is due to the fact
\begin{equation}
iH_{S}=i\sum_{j=1}^{N-1}\left( E_{1}+E_{2}+\cdots+E_{j}\right) h_{j},
\end{equation}

Recalling that we also have $iH_{A}^{0}\in\mathcal{L}$, we obtain
\begin{eqnarray}
\lefteqn{iH_{0}^{\prime\prime}\equiv i\left(
H_{0}-H_{S}\otimes1_{A}-H_{A}-H_{SA}\right)}  \notag \\
&& = i1_{S}\otimes\sum_{j=1}^{M-1}c_{j}\sigma_{x}^{j}\sigma_{x}^{j+1}+i\sum
_{j=1}^{N-1}d_{j}x_{j}\otimes1_{A}\in\mathcal{L}.  \label{32}
\end{eqnarray}

It then follows that
\begin{equation}
\left[ \left[ iH_{0}^{\prime \prime },iH_{y}^{1}\right] ,iH_{y}^{1}\right]
=-i4c_{1}\left( 1_{S}\otimes \sigma _{x}^{1}\sigma _{x}^{2}\right) \in
\mathcal{L},
\end{equation}%
yielding $i\left( 1_{S}\otimes \sigma _{x}^{1}\sigma _{x}^{2}\right) \in
\mathcal{L}$ thanks to the condition $c_{1}\neq 0$. This leads to the result
\begin{align*}
& \left[ \left[ iH_{0}^{\prime \prime }-ic_{1}1_{S}\otimes \sigma
_{x}^{1}\sigma _{x}^{2},\,iH_{y}^{2}\right] ,\,iH_{y}^{2}\right] \\
& =-i4c_{2}\left( 1_{S}\otimes \sigma _{x}^{2}\sigma _{x}^{3}\right) \in
\mathcal{L},
\end{align*}%
namely, $i\left( 1\otimes \sigma _{x}^{2}\sigma _{x}^{3}\right) \in \mathcal{%
L}$ since $c_{2}\neq 0$. Repeating this process we can finally prove
\begin{equation}
i\left( 1_{S}\otimes \sigma _{x}^{j}\sigma _{x}^{j+1}\right) \in \mathcal{L}%
,\ \ j=1,2,\cdots ,M-1.  \label{sigmaxxx}
\end{equation}%
Then the lemma follows from Lemma 2.

\begin{theorem}
\label{theorem1} When $M>2$, if Condition 1, Condition 2 and Condition 3 are
satisfied, then we have $\mathcal{L}=su(2^MN)$.
\end{theorem}

\noindent\textbf{Proof.} First we claim that under the conditions of the
theorem, for $j=1,2,\cdots,N-1$
\begin{equation}
i\left( x_{j}\otimes1_{A}\right), i\left( y_{j}\otimes1_{A}\right) \in%
\mathcal{L}.  \label{xyinlie}
\end{equation}
Recall that $iH_S\otimes 1_A\in\mathcal{L}$ and notice that Eq.(\ref%
{sigmaxxx}) implies $iH_{A}^{\prime}\in\mathcal{L}$, and hence
\begin{equation*}
iH_{S}^{\prime}=iH_{0}^{\prime\prime}-iH_{A}^{\prime}\in\mathcal{L}.
\end{equation*}
Then according to the result of Ref.[10,11], if Condition 3 is satisfied the
elements $i\left( x_{j}\otimes1_{A}\right)$ and $i\left(
y_{j}\otimes1_{A}\right)$ are contained in the subalgebra of $\mathcal{L}$
generated by $iH_S\otimes 1_A$ and $iH_{S}^{\prime}$. This proves the claim.

Since the elements of the set $\{ix_{jk},iy_{jk},ih_j|1\leq j<k\leq N\}$ can
be generated from the set $\{ix_j,iy_j|j=1,2,\cdots,N-1\}$ it follows from
Lemma 3, Lemma 4 and (\ref{xyinlie}) that the following elements are in the
Lie algebra $\mathcal{L}$
\begin{align*}
& ix_{jk}\otimes1_{A},\ \ y_{jk}\otimes1_{A},\ \ h_{j}\otimes1_{A}, \\
& x_{jk}\otimes\sigma_{[\alpha]},\ \ y_{jk}\otimes \sigma_{[\alpha]},\ \
h_{j}\otimes\sigma_{[\alpha]}, \\
& 1_{S}\otimes\sigma_{[\alpha]},
\end{align*}
where $[\alpha]\ne (0,0,\cdots,0)$, and $1\leq j<k\leq N$. It is easily
check that these elements are linearly independent and the total number of
these elements is
\begin{align}
\lefteqn{(N^{2}-1)+(N^{2}-1)(4^{M}-1)+(4^{M}-1)}  \notag \\
& =(2^{M}N)^{2}-1={\dim}\left( su(2^{M}N)\right).
\end{align}
This proves the theorem.

Before leaving this section we would like to note that the coupling between
the system and the accessor plays an essential role in the indirect control.
In the above given $H_{SA}$ there are $2(N-1)\times2^{M}$ coupling terms.
Actually as far as the controllability is concerned, we have simpler choices
of $H_{SA}$. For example, we can reduce the number of coupling terms to $%
2(N-1)$, just enough to guarantee the satisfaction of Condition 2.


\section{Indirect control for two-dimensional system}


In this section we will consider an explicit example, the indirect control
of a two-energy level system, to illustrate the general approach given in
last section. We also present a simpler indirect control scheme for
2-dimensional system.

\begin{figure}[tbp]
\includegraphics[bb=90 150 400 700,width=6 cm, clip]{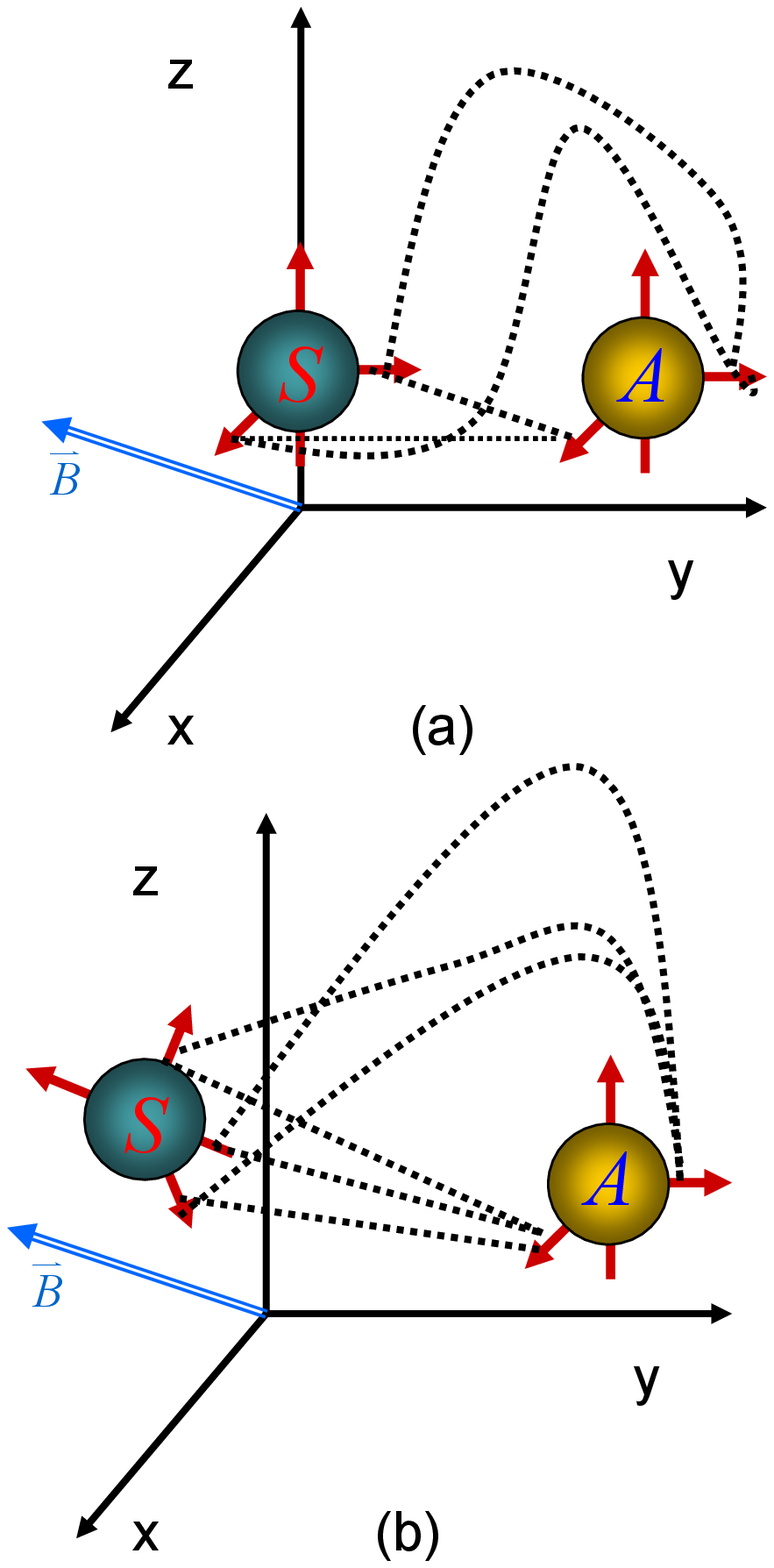}
\caption{(a) There are 4 terms (denoted by 4 dotted lines ) in the
interaction between the controlled qubit (green) and the quantum accessor
(yellow) when a constant field $\vec{B}$ is applied in the x-direction ; (b)
After the controlled qubit is rotated to be along the direction of the total
external field there will be 6 terms in the interaction, which are denoted
by 6 dotted lines.}
\label{fig:rotation}
\end{figure}

The 2-dimensional quantum system can be described by the Hamiltonian
\begin{equation}
H_{S}=\hbar\omega_{S}\sigma_{z}\otimes1_{A},
\end{equation}
in terms of Pauli's matrices. In this case, it is possible to use just one
qubit as the accessor. The Hamiltonian of the entire control system can be
written as
\begin{align}
H &
=\hbar\omega_{S}\sigma_{z}\otimes1_{A}+g\sigma_{x}\otimes1+1_{S}\otimes\hbar%
\omega_{I}\sigma_{z}  \notag \\
& +g_{xx}\sigma_{x}\otimes\sigma_{x}+g_{xy}\sigma_{x}\otimes\sigma _{y}
\notag \\
& +g_{yx}\sigma_{y}\otimes\sigma_{x}+g_{yy}\sigma_{y}\otimes\sigma _{y}
\notag \\
& +f_{1}(t)\left( 1_{S}\otimes\sigma_{x}\right) +f_{2}(t)\left(
1_{S}\otimes\sigma_{y}\right) .
\end{align}
Here we remark that the excitation term $\sigma_x\otimes 1$ can be removed
by rotating the controlled system around y-direction so that $%
\hbar\omega_{S}\sigma_{z}\otimes1_{A}+g\sigma_{x}\otimes1$ becomes $%
\hbar\omega_{S}^{\prime}\sigma_{z}\otimes1_{A}$ . As the price paid, the
rotated Hamiltonian contains the terms $g_{zx}^{\prime}\sigma_{z}\otimes%
\sigma_{x}$ and $g_{zy}^{\prime}\sigma _{z}\otimes\sigma_{y}$ (see Fig.%
\ref{fig:rotation}):
\begin{align}
H & =\hbar\omega_{S}^{\prime}\sigma_{z}\otimes1_{A}+1_{S}\otimes\hbar
\omega_{I}\sigma_{z}  \notag \\
& +g_{xx}^{\prime}\sigma_{x}\otimes\sigma_{x}+g_{xy}^{\prime}\sigma
_{x}\otimes\sigma_{y}  \notag \\
& +g_{yx}\sigma_{y}\otimes\sigma_{x}+g_{yy}\sigma_{y}\otimes\sigma _{y}
\notag \\
& +g_{zx}^{\prime}\sigma_{z}\otimes\sigma_{x}+g_{zy}^{\prime}\sigma
_{z}\otimes\sigma_{y}  \notag \\
& +f_{1}(t)\left( 1_{S}\otimes\sigma_{x}\right) +f_{2}(t)\left(
1_{S}\otimes\sigma_{y}\right) .  \label{remove}
\end{align}

The following theorem is the main result of this section.

\begin{theorem}
\label{theorem2} Suppose that $g_{xy}g_{yx} \neq g_{xx}g_{yy}$. Then the
symplectic Lie algebra \textrm{sp(4)} is included in $\mathcal{L}$.
Moreover, if $g\neq0$ is also satisfied, then $\mathcal{L}=su(4)$.
\end{theorem}

\noindent\textbf{Proof.} We observe that in the present case, Lemma 1
reduces to the trivially true identity since the coupling term in $H_A$ does
not appear. On the other hand the assumption $g_{xy}g_{yx} \neq g_{xx}g_{yy}$
simply means that Condition 2 is satisfied. Therefore Lemma 2 is valid.
Noticing that,by definition, $x_1=\sigma_x$, $y_1=\sigma_y$ and $%
h_1=\sigma_z $ with respect to a proper basis when $N=2$, we conclude, from
Lemma 2 and the fact that $\mathcal{L}$ contains the elements $%
i(1_S\otimes\sigma_x)$ $i(1_S\otimes\sigma_x)$ by definition, that $\mathcal{%
L}$ contains the following elements:

\begin{align}
& i(1_{S}\otimes\sigma_{\alpha}),\ \ \ \alpha=x,y,z;  \notag \\
& i\sigma_{\alpha}\otimes\sigma_{\beta},\ \ \ \alpha=x,y,\beta =x,y,z;
\notag \\
& i(\sigma_{z}\otimes1_{A})  \label{symgen}
\end{align}
and thus contains the element $g(i\sigma_x\otimes 1_A)$, which is obtained
by subtracting from $iH_0$ all the other terms, which lie in $\mathcal{L}$.

Now we claim that we can choose a basis of \textrm{sp(4)} from those
elements in (\ref{symgen}). In fact, we have
\begin{equation}
i\sigma_{z}\otimes1=\left(
\begin{tabular}{cc|cc}
i &  &  &  \\
& -i &  &  \\ \hline
&  & -i &  \\
&  &  & i%
\end{tabular}
\right) ,i(1\otimes\sigma_{z})=\left(
\begin{tabular}{cc|cc}
i &  &  &  \\
& i &  &  \\ \hline
&  & -i &  \\
&  &  & -i%
\end{tabular}
\right),
\end{equation}%
\begin{equation}
i\sigma_{x}\otimes\sigma_{z}=\left(
\begin{tabular}{cc|cc}
& i &  &  \\
i &  &  &  \\ \hline
&  &  & -i \\
&  & -i &
\end{tabular}
\right) ,i\sigma_{y}\otimes\sigma_{z}=\left(
\begin{tabular}{cc|cc}
& 1 &  &  \\
-1 &  &  &  \\ \hline
&  &  & -1 \\
&  & 1 &
\end{tabular}
\right),
\end{equation}%
\begin{equation}
i\sigma_{x}\otimes\sigma_{x}=\left(
\begin{tabular}{cc|cc}
&  & i &  \\
&  &  & i \\ \hline
i &  &  &  \\
& i &  &
\end{tabular}
\right), i\sigma_{x}\otimes\sigma_{y}=\left(
\begin{tabular}{cc|cc}
&  & 1 &  \\
&  &  & 1 \\ \hline
-1 &  &  &  \\
& -1 &  &
\end{tabular}
\right),
\end{equation}%
\begin{equation}
i\sigma_{y}\otimes\sigma_{x}=\left(
\begin{tabular}{cc|cc}
&  & 1 &  \\
&  &  & -1 \\ \hline
-1 &  &  &  \\
& 1 &  &
\end{tabular}
\right) , i\sigma_{y}\otimes\sigma_{y}=\left(
\begin{tabular}{cc|cc}
&  & -i &  \\
&  &  & i \\ \hline
-i &  &  &  \\
& i &  &
\end{tabular}
\right),
\end{equation}
\begin{equation}
i(1\otimes\sigma_{x})=\left(
\begin{tabular}{cc|cc}
&  &  & i \\
&  & i &  \\ \hline
& i &  &  \\
i &  &  &
\end{tabular}
\right) ,i(1\otimes\sigma_{y})=\left(
\begin{tabular}{cc|cc}
&  &  & 1 \\
&  & 1 &  \\ \hline
& -1 &  &  \\
-1 &  &  &
\end{tabular}
\right),  \label{matrixelement}
\end{equation}
with respect to the ordered basis $\{|0\rangle\otimes
|0\rangle,|1\rangle\otimes |0\rangle,|1\rangle\otimes
|1\rangle,|0\rangle\otimes |1\rangle\}$. It is readily check that these
matrices are linearly independent and satisfy the equation
\begin{equation}
S^{t}x+xS=0,
\end{equation}
the defining relation of $sp(4)$, where
\begin{equation}
S=\left(
\begin{array}{cc}
& I \\
-I &
\end{array}
\right)
\end{equation}
and $I$ is the $2\times2$ identity matrix. This proves the claim, and hence
the first part of the theorem, as the dimension of $sp(4)$ is $10$.

If $g\ne 0$, from $g(i\sigma_x\otimes 1_A)\in \mathcal{L}$ we can derive $%
i\sigma_x\otimes 1_A\in \mathcal{L}$. It is easily check that this element,
together with the elements in (\ref{symgen}), can generate $15$ linearly
independent elements by Lie bracket operations. As the dimension of $su(4)$
is exactly $15$ we conclude that $\mathcal{L}=su(4)$. The proof of Theorem 2
is thus completed.

We remark that it is easy to satisfy the condition $g_{xy}g_{yx} \neq
g_{xx}g_{yy}$. For example, we can take
\begin{equation}
g_{xx} = g_{yy} = 0,g_{xy} = g_{yx} \ne 0,
\end{equation}
or
\begin{equation}
g_{xy} = g_{yx} = 0,g_{xx} = g_{yy} \ne 0.
\end{equation}
In both cases, there are only two terms in the coupling between the system $%
\mathcal{S}$ and the accessor $\mathcal{A}$.

Finally, we point out that, by making full use of the property that the
square of Pauli's matrices is unity, which is peculiar to the $N=2$ case, we
can manage to control the system completely by means of simpler couplings
between the system and the accessor. Let us consider, as an example, the
control system
\begin{align}
H_{0} & =\hbar\omega_{S}\sigma_{z}\otimes1_{A}+g\sigma_{x}\otimes1  \notag \\
& +1_{S}\otimes\hbar\omega_{I}\sigma_{z}+g_{xx}\sigma_{x}\otimes\sigma_{x},
\\
H_{c} & =f_{1}(t)\left( 1_{S}\otimes\sigma_{x}\right) +f_{2}(t)\left(
1_{S}\otimes\sigma_{y}\right)  \notag
\end{align}
where $g\neq0$ and $g_{xx}\neq0$. Such a control system is essentially
different from the system just discussed above as in this case Condition 2
is never satisfied. One can easily check that
\begin{align}
& (2g_{xx})^{-1}\left( -[iH_{0},i(1\otimes\sigma_{y})]+2i\hbar\omega
_{I}\otimes\sigma_{x}\right)  \notag \\
& =i\sigma_{x}\otimes\sigma_{z}\in\mathcal{L},  \label{2w1}
\end{align}
from which we further have
\begin{align}
& -(2\hbar\omega_{S})^{-1}\left[ iH_{0}-\hbar\omega_{I}1\otimes\sigma
_{z}-ig_{xx}\sigma_{x}\sigma_{x},\ i\sigma_{x}\otimes\sigma_{z}\right]
\notag \\
& =(2\hbar\omega_{S})^{-1}\left[ \hbar\omega_{S}\sigma_{z}\otimes
1_{A}+g\sigma_{x}\otimes1,\ \sigma_{x}\otimes\sigma_{z}\right]  \notag \\
& =i\sigma_{y}\otimes\sigma_{z}\in\mathcal{L}.  \label{2w2}
\end{align}
Now it should not be difficult to proceed further to prove that the two
conclusions of Theorem 2 are still valid though the premise is no longer
true. We leave the details to interested readers.

\section{Indirect control for 3-dimensional quantum system}


In this section we discuss the indirect control of 3-dimensional quantum
system based on the approach presented in Section III.

Since Theorem 1 is, generally speaking, not valid when $M\leq 2$, we first
consider the possibility of using 3 qubits to control the system, namely, we
assume that $M=3$.

Let $[\beta]_1=(x,x,x)$, $[\beta]_2=(x,x,y)$, $[\beta]_3=(x,y,x)$ and $%
[\beta]_4=(y,x,x)$. To satisfy Condition 2, we can simply choose $%
g_{[\beta]}^{j(k)}=0$ except that
\begin{equation}
g_{[\beta]_1}^{1(1)}=g_{[\beta]_2}^{2(1)}=g_{[\beta]_3}^{1(2)}
=g_{[\beta]_4}^{2(2)}=1,  \label{3dc0}
\end{equation}
namely,
\begin{align}
H_{SA} &
=x_{1}\otimes\sigma_{y}^{1}\sigma_{y}^{2}\sigma_{y}^{3}+x_{2}\otimes%
\sigma_{y}^{1}\sigma_{y}^{2}\sigma_{x}^{3}  \notag \\
& +y_{1}\otimes\sigma_{y}^{1}\sigma_{x}^{2}\sigma_{y}^{3}+y_{2}\otimes
\sigma_{x}^{1}\sigma_{y}^{2}\sigma_{y}^{3}.
\end{align}

In fact, in such a case, we have
\begin{equation}
\det\left[
\begin{array}{cccc}
g_{[\beta]_1}^{1(1)} & g_{[\beta]_1}^{2(1)} & g_{[\beta]_1}^{1(2)} &
g_{[\beta]_1}^{2(2)} \\
g_{[\beta]_2}^{1(1)} & g_{[\beta]_2}^{2(1)} & g_{[\beta]_2}^{1(2)} &
g_{[\beta]_2}^{2(2)} \\
g_{[\beta]_3}^{1(1)} & g_{[\beta]_3}^{2(1)} & g_{[\beta]_3}^{1(2)} &
g_{[\beta]_3}^{2(2)} \\
g_{[\beta]_4}^{1(1)} & g_{[\beta]_4}^{2(1)} & g_{[\beta]_4}^{1(2)} &
g_{[\beta]_4}^{2(2)}%
\end{array}
\right]=1  \label{3matrix}
\end{equation}

Now assume Condition 1, then Condition 3 is enough to guarantee the complete
controllability. In our present case, Condition 3 has a simple form\cite%
{fu1,fu2}:
\begin{equation}
\Delta_{21}^{2}\neq\Delta_{32}^{2}\text{ and }d_{1}\neq0,d_{2}\neq 0
\label{con31}
\end{equation}

or
\begin{equation}
\Delta_{21}^{2}=\Delta_{32}^{2}\text{ and }d_{1}\neq\pm d_{2}\neq0,
\label{con32}
\end{equation}

where $\Delta_{jk}\equiv E_j-E_k$ ($3\geq j>k\geq 1$) is the energy gap.
\bigskip

Now we consider the possibility of using only two qubits to control the
3-dimensional system. As in this case $M=2$, the general approach developed
in Section III cannot be fully applied. However, we have the following
conclusion: if we can control not only each qubit, but also their coupling
independently, we can indirectly control the 3-dimensional system using two
qubits. In fact, if this is the case, we can take the Hamiltonian as
\begin{align}
H & =H_{0}+H_{c}^{1}+H_{c}^{2}+H_{c}^{12}  \notag \\
H_{0} & =\sum_{j=1}^{3}\hbar\omega_{S}e_{jj}\otimes1_{A}+\left(
d_{1}x_{1}+d_{2}x_{2}\right) \otimes1_{A}  \notag \\
& +1_{S}\otimes\sum_{j=1}^{2}(\hbar\omega_{I}\sigma_{z}^{j}) \\
&
+\sum_{j=1}^{2}\sum_{\alpha_{1},\alpha_{2}=x,y}g_{\alpha_1%
\alpha_2}^{j(k)}s_{j}^{(k)}\otimes(\sigma_{\alpha_{1}}^{1}\sigma_{%
\alpha_{2}}^{2})  \notag \\
H_{c}^{j} & =f_{j}(t)\left( 1_{S}\otimes\sigma_{x}^{j}\right)
+f_{j}^{\prime}(t)\left( 1_{S}\otimes\sigma_{y}^{j}\right)  \notag \\
H_{c}^{12} & =f(t)1_{S}\otimes\sigma_{x}^{1}\sigma_{x}^{2}.
\end{align}
Let $\mathcal{L}$ be the Lie algebra generated by the elements
\begin{equation}
iH_{0},\ \ \ i\left( 1_{S}\otimes\sigma_{x}^{j}\right) ,\ \ \ i\left(
1_{S}\otimes\sigma_{y}^{j}\right) ,\ \ \ i\left(
1\otimes\sigma_{x}^{1}\sigma_{x}^{2}\right) ,
\end{equation}
where $j=1,2$. Then mathematically the complete controllability condition is
$\mathcal{L}=su(4)$. Using a method similar to that in Section III we can
prove $\mathcal{L}=su(4)$ if the condition(\ref{con31}) or (\ref{con32}),
and the condition
\begin{equation}
\det\left[
\begin{array}{cccc}
g_{xx}^{1(1)} & g_{xx}^{2(1)} & g_{xx}^{1(2)} & g_{xx}^{2(2)} \\
g_{xy}^{1(1)} & g_{xy}^{2(1)} & g_{xy}^{1(2)} & g_{xy}^{2(2)} \\
g_{yx}^{1(1)} & g_{yx}^{2(1)} & g_{yx}^{1(2)} & g_{yx}^{2(2)} \\
g_{yy}^{1(1)} & g_{yy}^{2(1)} & g_{yy}^{1(2)} & g_{yy}^{2(2)}%
\end{array}
\right] \neq0  \label{con3}
\end{equation}
are satisfied. We would rather omit the details to avoid redundancy.

Finally, we conclude this section by pointing out that (\ref{con3}) can be
satisfied by simply choosing
\begin{align}
H_{SA}^{\prime} &
=x_{1}\otimes\sigma_{x}^{1}\sigma_{x}^{2}+y_{1}\otimes\sigma_{x}^{1}%
\sigma_{y}^{2}  \notag \\
&
+x_{2}\otimes\sigma_{y}^{1}\sigma_{x}^{2}+y_{2}\otimes\sigma_{y}^{1}%
\sigma_{y}^{2}.
\end{align}


\section{Conclusion and remarks}


In this paper we investigated the controllability of an arbitrary finite
dimensional quantum system via a quantum accessor modeled as a spin chain
with nearest neighbor coupling of XY-type. The general approach is applied
to the indirect control of two and three dimensional quantum systems. We
also present indirect control schemes simpler than the general scheme for
two and three dimensional systems. Our approach shows that one can
completely control an finite-dimensional quantum system through a quantum
accessor if the system and the accessor are coupled properly.

We point out that we have supposed that each spin of the quantum accessor
can be individually controlled. In forthcoming paper we would like to
explore the indirect control of the quantum systems by controlling the
accessor globally. Global control of spin chains itself has been studied
recently in the context of quantum computation \cite{global}. It is
definitely of interest to realize the indirect control by global control of
quantum accessor. In Sec.\thinspace IV we found that we can achieve the
indirect control without applying the constant excitation field to the system by
rotating the system around y-direction (see Eq.\thinspace (\ref{remove})).
This example suggests us removing the excitation field from the controlled
system to achieve the pure indirect control. We will address this issue in
our forthcoming paper. Obviously it is also significant study a control
system where the fixed interaction between the controlled system and the
accessor is so weak that it can be neglected approximately when the strong
field, which controls the accessor, is switched on.

Before concluding this paper we would like to remark that in the
conventional investigation on the controllability of quantum systems, the
controls are usually classical or semiclassical since the controlling field
is described as a time-dependent functions and directly affects the time
evolution of the closed or open quantum systems to be controlled~\cite%
{albe,viol1,viol2,alta,Rama}. So it might be more appropriate to name those
types of control \emph{(semi)classical control of quantum systems}.

\section*{Acknowledgement}

This work is supported by the NSFC with grant No.\thinspace 10675085,
90203018, 10474104 and 60433050, \ and NFRPC with No.\thinspace 2006CB921205
and 2005CB724508.

\end{document}